\documentclass[useAMS,usenatbib]{mn2e}

\usepackage{graphicx}
\usepackage{amsmath}
\usepackage[dvips,usenames]{color}
\def\aj{AJ}%
\def\apj{ApJ}%
\def\apjl{ApJ}%
\def\aap{A\&A}%
\def\aaps{A\&AS}%
\def\mnras{MNRAS}%
\def\pasp{PASP}%
\usepackage{times}

\newcommand{\mini}{\mbox{$M_{\rm i}$}}

\newcommand{\thef}{\mbox{$t_{\rm HeF}$}}
\newcommand{\Mhef}{\mbox{$M_{\rm HeF}$}}

\newcommand{\Msun}{\mbox{$M_{\odot}$}}
\newcommand{\Teff}{\mbox{$T_{\rm eff}$}}

\newcommand{\comment}[1]{}
\newcommand{\beq}{\begin{equation}}
\newcommand{\eeq}{\end{equation}}
\newcommand{\beqa}{\begin{eqnarray}}
\newcommand{\eeqa}{\end{eqnarray}}

\title[Two red clumps in NGC~419]
{Discovery of two distinct red clumps in NGC~419: \\ 
a rare snapshot of a cluster at the onset of degeneracy}

\author[Girardi, Rubele \& Kerber]{L\'eo Girardi$^{1}$, Stefano Rubele$^{1,2}$
	and Leandro Kerber$^{3}$\\
$^{1}$ Osservatorio Astronomico di Padova -- INAF,
	Vicolo dell'Osservatorio 5, I-35122 Padova, Italy \\
$^{2}$ Dipartimento di Astronomia, Universit\`a di Padova,
	Vicolo dell'Osservatorio 2, I-35122 Padova, Italy \\
$^{3}$ Universidade de S\~ao Paulo, IAG, Rua do Mat\~ao 1226, 
	Cidade Universit\'aria, S\~ao Paulo 05508-900, Brazil
}

\begin{document}

\date{{\bf To appear in MNRAS Letters (www.blackwell-synergy.com).} 
Accepted 2008 Dec. 10. Received 2008 Dec. 9;  
in original form 2008 Nov. 14}

\pagerange{\pageref{firstpage}--\pageref{lastpage}} \pubyear{2008}

\maketitle

\label{firstpage}

\begin{abstract}
Colour--magnitude diagrams (CMD) of the SMC star cluster NGC~419,
derived from HST/ACS data, reveal a well-delineated secondary clump
located below the classical compact red clump typical of
intermediate-age populations. We demonstrate that this feature belongs
to the cluster itself, rather than to the underlying SMC field. Then,
we use synthetic CMDs to show that it corresponds very well to the
secondary clump predicted to appear as a result of He-ignition in
stars just massive enough to avoid e$^-$-degeneracy settling in their
H-exhausted cores. The main red clump instead is made of the slightly
less massive stars which passed through e$^-$-degeneracy and ignited
He at the tip of the RGB. In other words, NGC~419 is the rare snapshot
of a cluster while undergoing the fast transition from classical to
degenerate H-exhausted cores. At this particular moment of a cluster's
life, the colour distance between the main sequence turn-off and the
red clump(s) depends sensitively on the amount of convective core
overshooting, $\Lambda_{\rm c}$. By coupling measurements of this
colour separation with fits to the red clump morphology, we are able
to estimate simultaneously the cluster mean age
($1.35_{-0.04}^{+0.11}$~Gyr) and overshooting efficiency
($\Lambda_{\rm c}\!=\!0.47_{-0.04}^{+0.14}$). Therefore, clusters like
NGC~419 may constitute important marks in the age scale of
intermediate-age populations.  After eye inspection of other CMDs
derived from HST/ACS data, we suggest that the same secondary clump
may also be present in the LMC clusters NGC~1751, 1783, 1806, 1846,
1852, and 1917.
\end{abstract}


\section{Introduction}
\label{intro}

In the last decade, wide field imagers and the Hubble Space Telescope
(HST) have provided detailed CMDs for the star fields in the
Magellanic Clouds. One of the main surprises was the discovery that
the red clump of core-He burners is not a compact feature, but may
present extensions amounting to a few 0.1~mag departing from its top
and bottom parts, as well as a blue extension that connects with the
horizontal branch of the old metal-poor populations.

The ground-based observations from \citet{Bica_etal98} and
\citet{Piatti_etal99} evinced a $\sim\!0.4$~mag extension of the red
clump to fainter magnitudes, spread over wide areas of the outer LMC
disk. \citet{Girardi99} gave a clear interpretation to this extention
-- thereafter named {\em secondary red clump} -- claiming that it is
made of the stars just massive enough to start burning He in
non-degenerate conditions, at ages of $\sim\!1$~Gyr, whereas the main
body of the red clump is made of all the intermediate-age and old
stars which passed through degenerate cores and the He-core flash at
the tip of the RGB. The same feature was suggested to be present in
the Hipparcos CMD for the Solar Neighbourhood
\citep{Girardi_etal98}, and provides an explanation to the vertical
extension -- amounting to about $0.8$~mag in the $F814W$ band
\citep[see e.g.][]{Holtzman_etal97} -- of the red clump in the sharp
CMDs obtained by the HST for several LMC fields.

Plots of the red clump magnitude versus age for star clusters -- as in
\citet[][figs. 3 and 4]{Girardi99}, and 
\citet[][fig. 6]{GrocholskiSarajedini02} -- seem to confirm the 
theoretical framework that led to \citeauthor{Girardi99}'s
(\citeyear{Girardi99}) prediction of a secondary clump: they clearly
indicate a red clump decreasing 
in luminosity up to $\sim\!1$~Gyr, then a jump upwards by about
0.4~mag, and a much slower evolution thereafter.

A handful of faint red clump stars are also found among the radial velocity
members of the open clusters NGC~752 and 7789, and possibly also in
NGC~2660 and 2204 \citep{Mermilliod_etal98,
GirardiMermilliod_etal00}. The latter authours associated these stars
with the secondary clump feature, and tried to figure out how it could
appear in an object for which the age spread was supposedly very
small. Indeed, from their discussion it is clear that a single cluster
age corresponds to a very narrow range of red clump masses and hence
to the sampling of either the faint (secondary) or the bright
(classical) red clump. The two red clumps could not appear together in
a single star cluster, unless some other mechanism -- e.g. a
dispersion in the mass loss along the RGB, or in the efficiency of
overshooting on the main sequence (MS) -- were invoked.

The reasoning from \citet{GirardiMermilliod_etal00} would however now
fail, at least for the star clusters in the Magellanic Clouds. Indeed,
now {\em we know} that many of them do not represent single ages, but
rather a range of ages that can extend up to a few 100~Myr
\citep{Mackey_etal08, Milone_etal08} -- or equivalently, to a range of
MS turn-off (MSTO) masses of a few 0.05~\Msun.

By visual inspection of the CMDs for a few SMC star clusters studied
by \citet{Glatt_etal08}, we came across what was {\em very evidently}
a composite structure of main+secondary red clump in NGC~419. The
presence of a secondary clump in this cluster was indeed noticed by
\citet[][their sect. 3.6]{Glatt_etal08}, who however have
suggested it to be ``a red clump of the old SMC field star
population''. They also noticed the composite structure in the MS,
attributing it to an extended period of star formation, similarly to
the one claimed by \citet{Mackey_etal08} for three LMC clusters.

In the following, we demonstrate that the composite structure of the
red clump in NGC~419 is real and undubiously associated to the cluster
(Sect.~\ref{data}). We show that it corresponds to the simultaneous
presence of stars which have started burning He under non-degenerate
and degenerate conditions (Sect.~\ref{models}). We then illustrate
that this rare occurrence in a cluster allows us to set stringent
constraints on the cluster age and amount of convective core
overshoooting during the MS evolution (Sect.~\ref{constraints}).
We then briefly suggest that NGC~419 is not a unique case, and draw
some final comments in Sect.~\ref{implications}.

\section{NGC~419 photometry and CMD}
\label{data}

We have retrieved from the HST archive
the NGC~419 data obtained by GO-10396 (PI: J.S. Gallagher).  The
dataset consists of a 740~arcsec$^2$ area observed with the Advanced
Camera for Surveys (ACS) High Resolution Channel (HRC) centered on
NGC~419, plus a $4.24\times10^4$~arcsec$^2$ area observed with the ACS
Wide Field Channel (WFC) 37\arcsec\ offset from the cluster centre.
Both datasets were reduced via standard procedures
\citep{Sirianni_etal05}. The HRC data, given its high spatial
resolution, is the most useful for the study of the NGC~419
population, whereas the WFC dataset provides the comparison data for
interpreting the SMC field.

We have performed aperture and PSF photometry on the calibrated HRC
image, finding that both provide
CMDs very similar to the \citet{Glatt_etal08} ones. We have then opted
for the PSF catalogue and cutted it at $( {\rm sharp}_{F555W}^2+{\rm
sharp}_{F814W}^2 )^{1/2}<0.2$.
This quality cut eliminates many outliers from the CMD, especially at
the faintest magnitudes, but do not affect the morphology of 
features at the red clump and MSTO level.

\begin{figure}
\resizebox{\hsize}{!}{\includegraphics{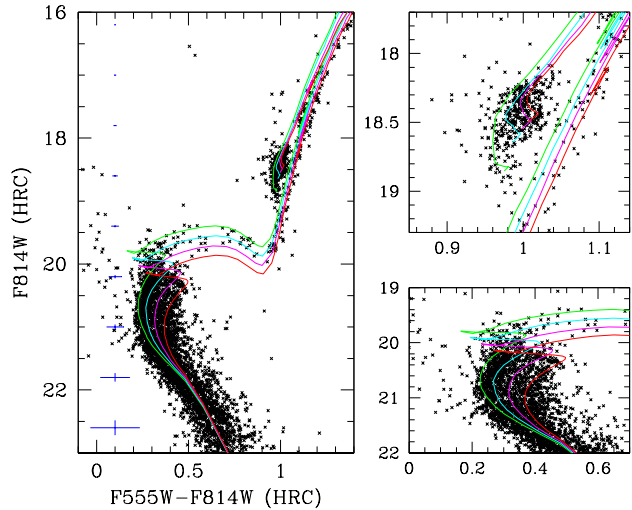}}
\caption{The CMD for NGC~419 as derived from the HRC data centered on 
the cluster (left panel). The $1\sigma$ error bars, as derived from
artificial star tests, are drawn at the left.
The right panels detail the red clump (top) and MSTO regions
(bottom). The overlaid isochrones are from \citet{Marigo_etal08}, for
a metallicity $Z\!=\!0.004$, ages varying from $\log(t/{\rm
yr})\!=\!9.10$ to $9.25$ with a constant spacing of 0.05~dex,
$E_{F555W\!-\!F814W}\!=\!0.09$, and
$(m\!-\!M)_{F814W}\!=\!18.85$. Notice that these particular isochrones
describe reverse sequences in the MSTO and red clump regions of the
CMD: whereas the MSTO gets dimmer for increasing ages, the red clump
gets brighter.}
\label{fig_cmd_hrc}
\end{figure}
\begin{figure}
\resizebox{\hsize}{!}{\includegraphics{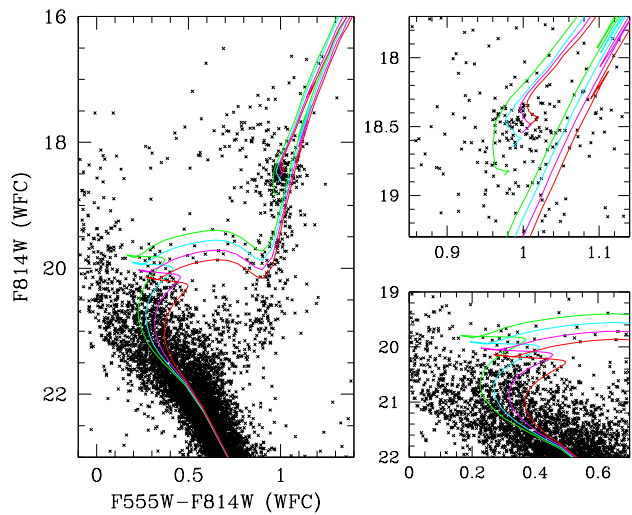}}
\caption{The CMDs for $2.47\times10^4$~arcsec$^2$ field around NGC~419, 
as derived from the WFC data after subtracting a circular area of
radius 75\arcsec\ around the cluster. The overlaid isochrones are the
same as in Fig.~\ref{fig_cmd_hrc}, and are plotted for reference
only. The bulk of the red clump is below the saturation limit at
$F555W\!\sim\!17.9$.}
\label{fig_cmd_wfc}
\end{figure}

The HRC data is plotted in Fig.~\ref{fig_cmd_hrc}, which shows both
the global CMD and separate panels detailing the red clump and the
MSTO regions. The extended nature of the red clump is evident in the
figure\footnote{The choice of the $F814W$ band for the ordinate is not
casual, since $F814W$ presents flatter bolometric corrections than
$F555W$, over the \Teff\ range of the red clump. This improves the
separation in magnitude of the clump substructures, and at the same
time keeps the subgiant branch out of its magnitude range.}. It seems
to be formed by a main blob located between $F814W\!=\!18.1$ and
18.65, followed by a well-defined faint wing between $F814W\!=\!18.65$
and 19.0, which we tentatively identify as the secondary clump. We
note that there is no similar wing extending from the top of the red
clump. These sequences are about 0.08~mag bluer (in $F555W\!-\!F814W$)
than the ridgeline of RGB stars. In addition, the red part of the CMD
shows a well delineated subgiant branch, and the bump of early
asymptotic giant branch (AGB) stars centered at $F814W\!=\!17.7$.

There are 55 stars between $F814W\!=\!18.65$ and
19.0, 8 out of which are red enough to belong to the RGB rather than
to the secondary clump. The main red clump instead contains 341
stars. These numbers correspond to the HRC effective area of
740~arcsec$^2$, centered on NGC~419. We verified that both sets of
stars distribute all over the HRC image, and share similar
distributions of photometric errors and sharpness. So, it is unlikely
that the secondary clump could be an artifact of a given subsample of
the data.

Can the field SMC stars account for the 47 stars in the secondary
clump, as suggested by \citet{Glatt_etal08}? To answer this question,
we look at the WFC data of Fig.~\ref{fig_cmd_wfc}, which covers an
area 33 times larger than the HRC one. It was obtained subtracting
from the complete WFC catalogue -- without applying any quality cut --
a circular area of 75\arcsec\ in radius around NGC~419. The remaining
area of $2.47\times10^4$~arcsec$^2$ was considered to be ``SMC
field'', and contains 150 red clump stars in the $F814W$ range between
18.1 and 19.0. Therefore, the expected number of these stars in the
740~arcsec$^2$ area of HRC is of just 4.5. Moreover, their typical
magnitudes are closer to those of the main red clump in NGC~419,
rather than to the secondary one. We conclude that the field cannot
contribute with more than $\sim$10\% of the 47 stars observed in the
secondary clump, and probably contribute much less. {\em The bulk of
the secondary clump observed in HRC data indeed belongs to NGC~419}.

\section{Modelling the two clumps}
\label{models}

Our interpretation of the red clump structure in NGC~419 is already
clear in Fig.~\ref{fig_cmd_hrc} and in the discussion of
Sect.~\ref{intro}: the fainter secondary red clump is explained by the
core-He burning stars belonging to the younger isochrones, which have
just avoided e$^-$-degeneracy before igniting He. This feature has
been throughfully discussed in \citet{Girardi99} and
\citet{Girardi_etal98}, in the context of galaxy field populations. It
appears naturally in simulations of star-forming galaxies with
moderate-to-high metallicities, provided that the underlying stellar
models do present a fine resolution in mass
\citep[better than 0.1~\Msun; see][]{Girardi99}.

In the following, we discuss the specific case of NGC~419 by means of
newly-computed evolutionary tracks of initial composition
$(Z\!=\!0.004, Y\!=\!0.250)$. The input physics is the same as in
\citet{Bertelli_etal08}. We have initially adopted a moderate amount
of convective overshooting, i.e. $\Lambda_{\rm c}\!=\!0.5$, where
$\Lambda_{\rm c}$ is the size of the overshooting region across the
convective boundary, in pressure scale heigths, following the
\citet{Bressan_etal81, Bressan_etal93} definitions. For a limited 
interval of initial masses \mini, typically going from
$\Mhef-0.4\,\Msun$ to $\Mhef+0.4\,\Msun$, we follow the
evolution up to the thermally pulsing AGB, whereas for smaller masses
(down to 0.6~\Msun) we have computed only the MS evolution. Stellar
tracks are spaced by $\Delta\mini\!=\!0.05$~\Msun. These tracks are
converted to stellar isochrones in the ACS/HRC and ACS/WFC Vegamag
systems using the transformations from \citet{Girardi_etal08}.

\begin{figure*}
\begin{minipage}{0.163\textwidth}
\resizebox{\hsize}{!}{\includegraphics{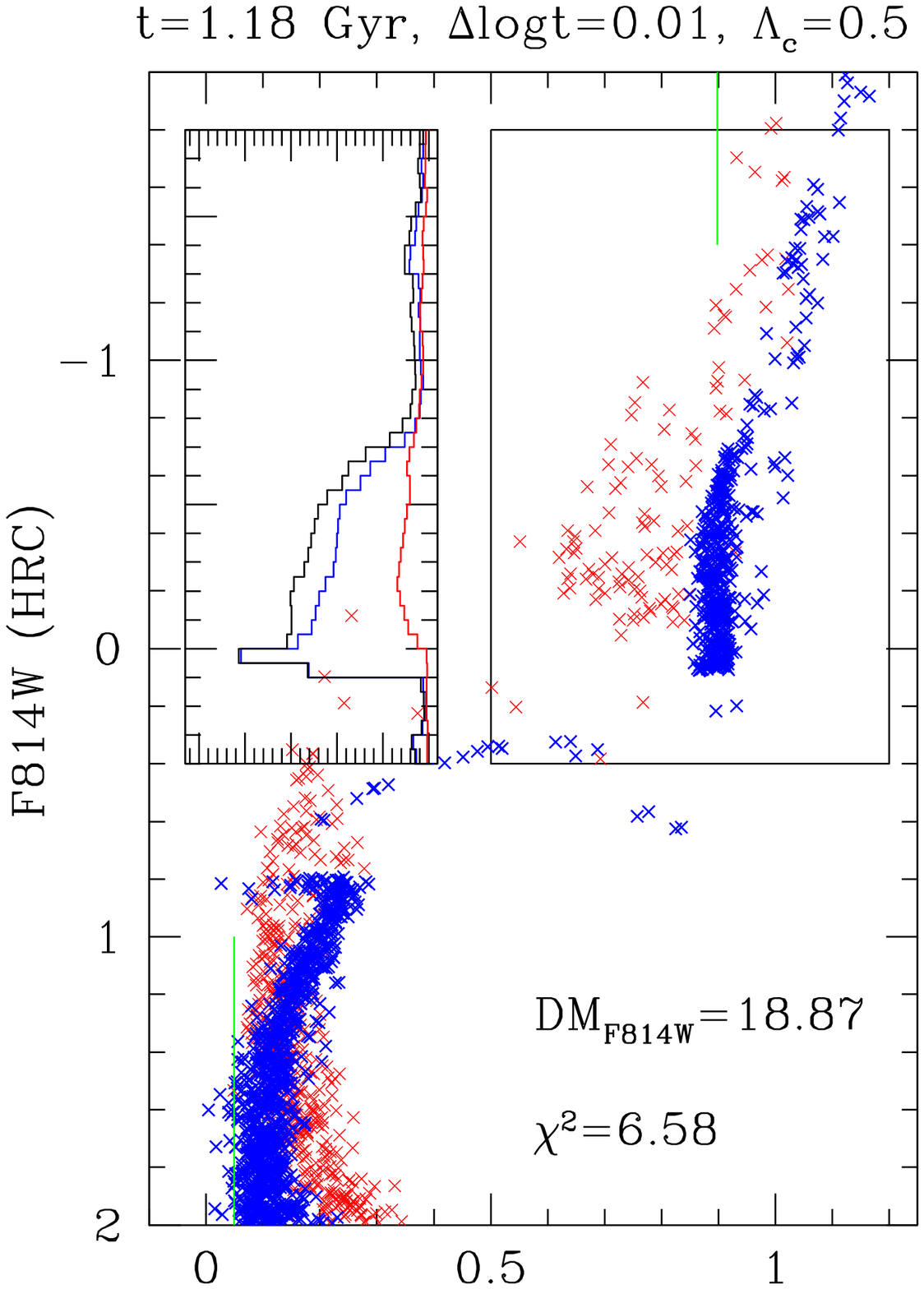}}
\end{minipage}
\begin{minipage}{0.163\textwidth}
\resizebox{\hsize}{!}{\includegraphics{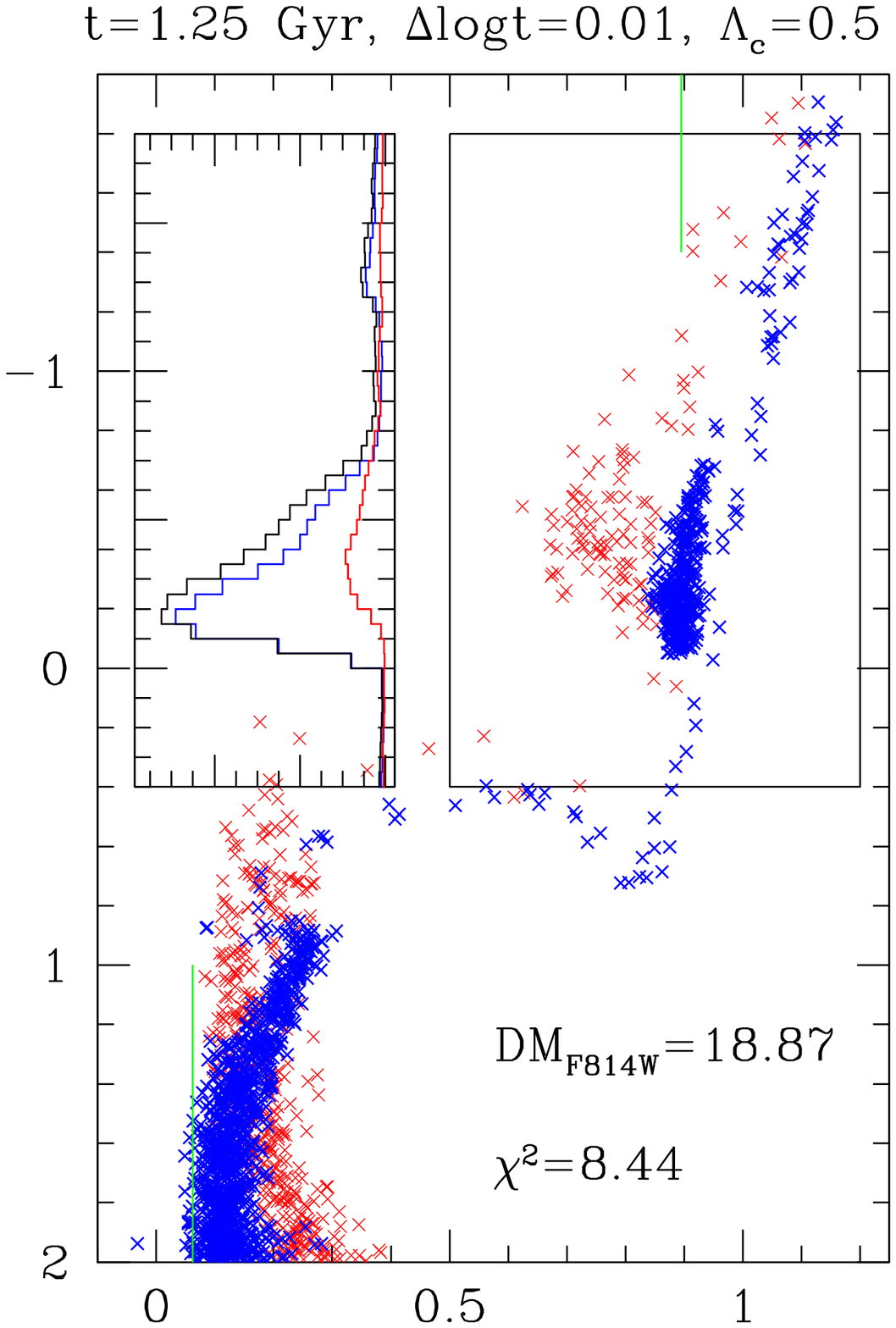}}
\end{minipage}
\begin{minipage}{0.163\textwidth}
\resizebox{\hsize}{!}{\includegraphics{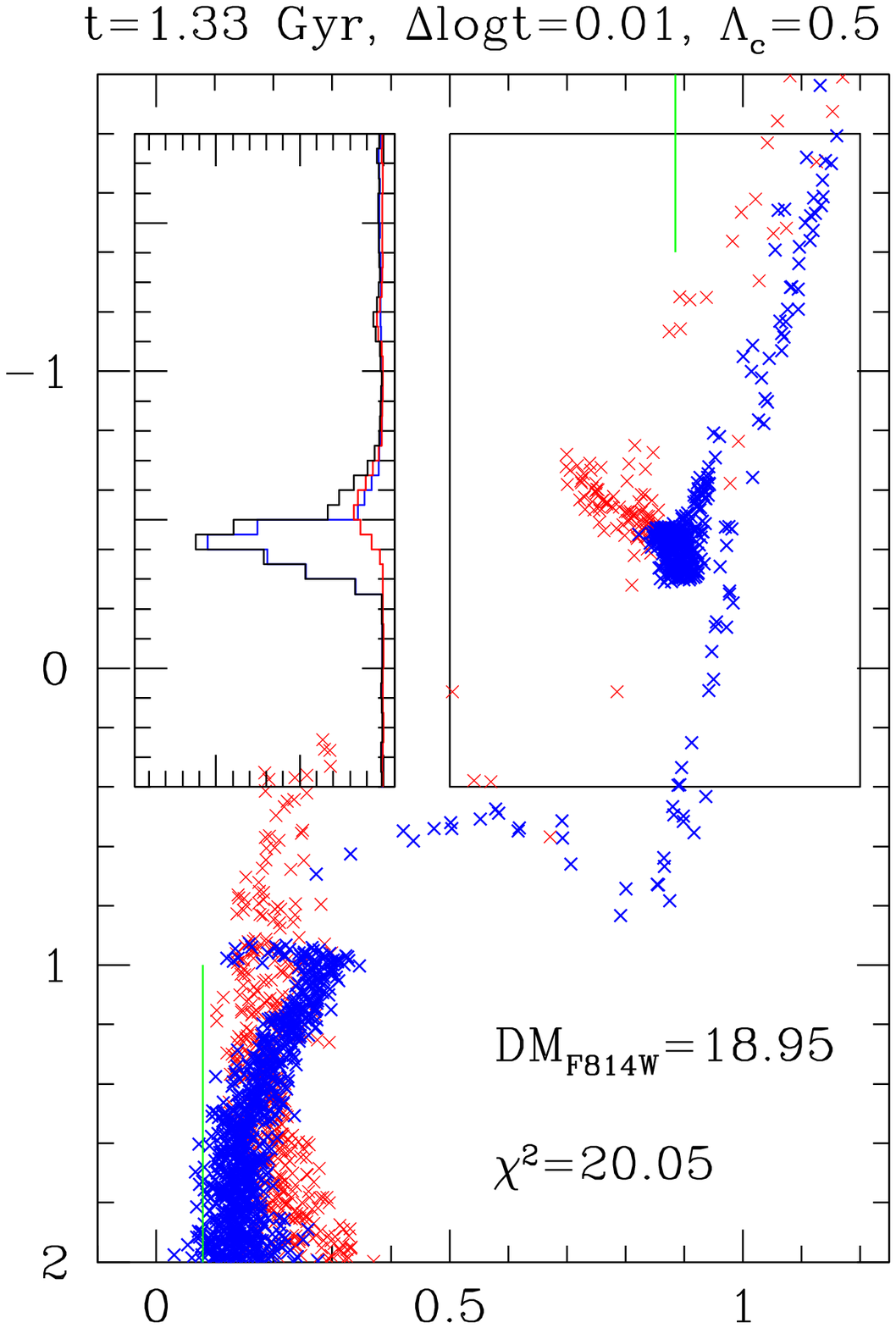}}
\end{minipage}
\begin{minipage}{0.163\textwidth}
\resizebox{\hsize}{!}{\includegraphics{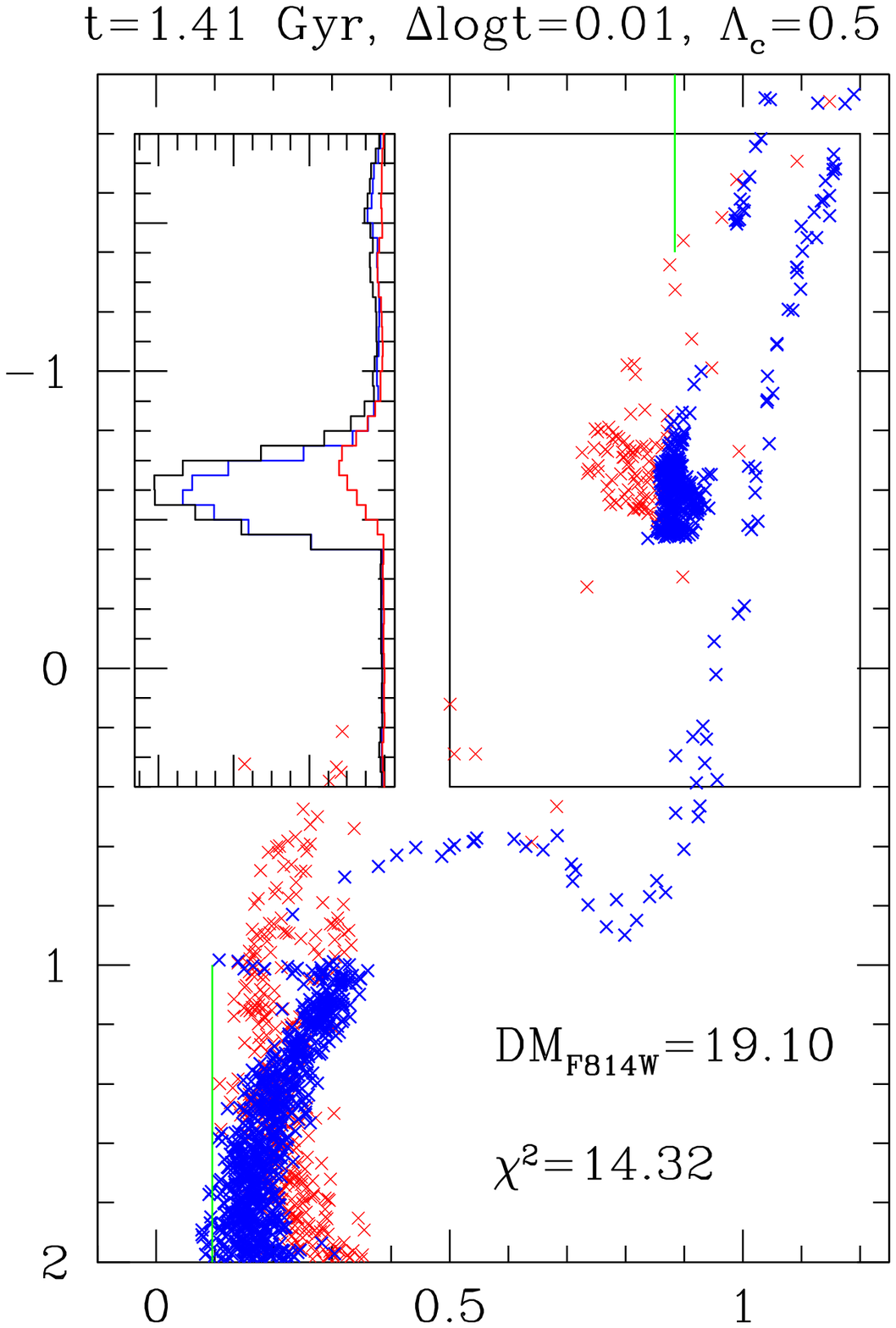}}
\end{minipage}
\begin{minipage}{0.163\textwidth}
\resizebox{\hsize}{!}{\includegraphics{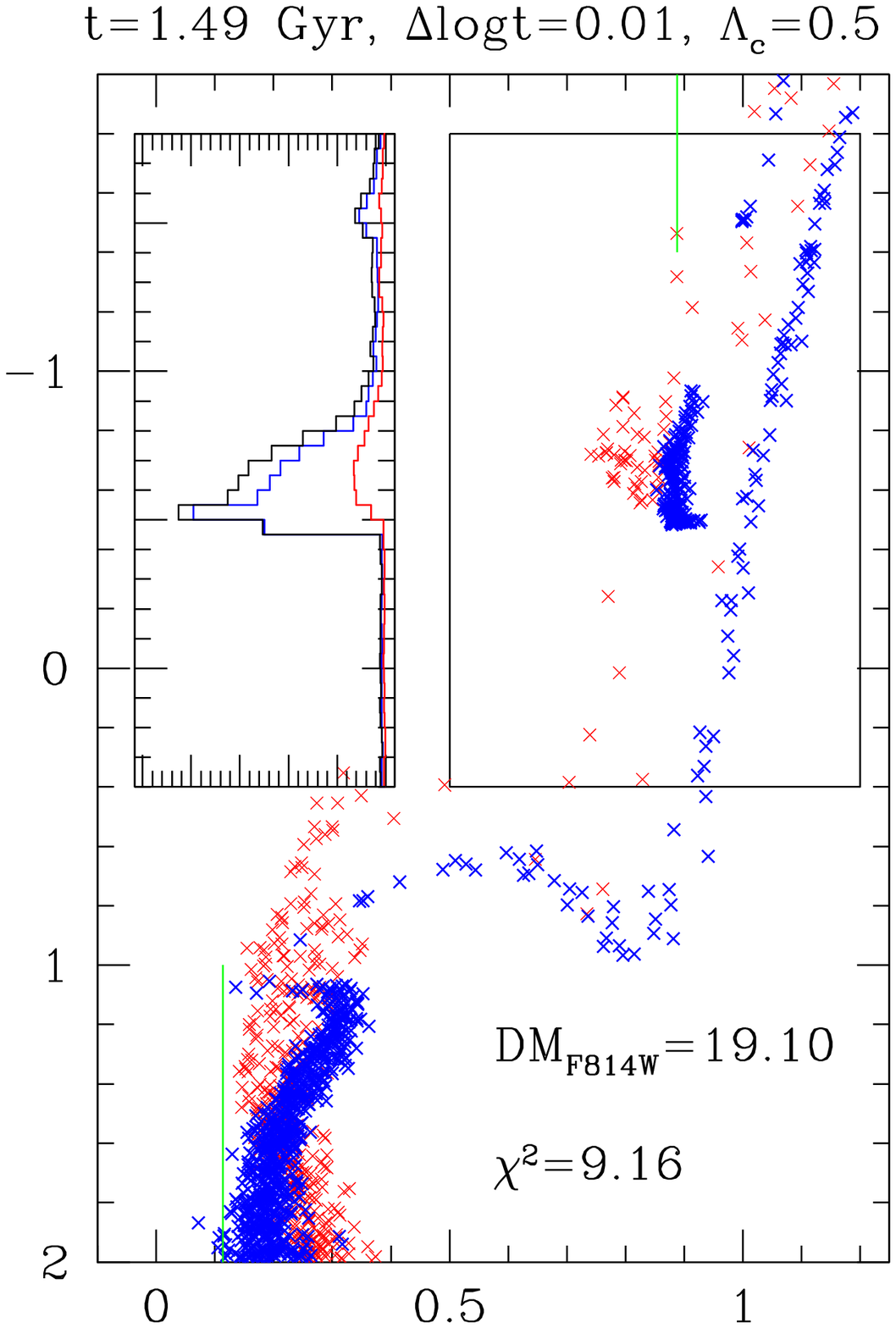}}
\end{minipage}
\begin{minipage}{0.163\textwidth}
\resizebox{\hsize}{!}{\includegraphics{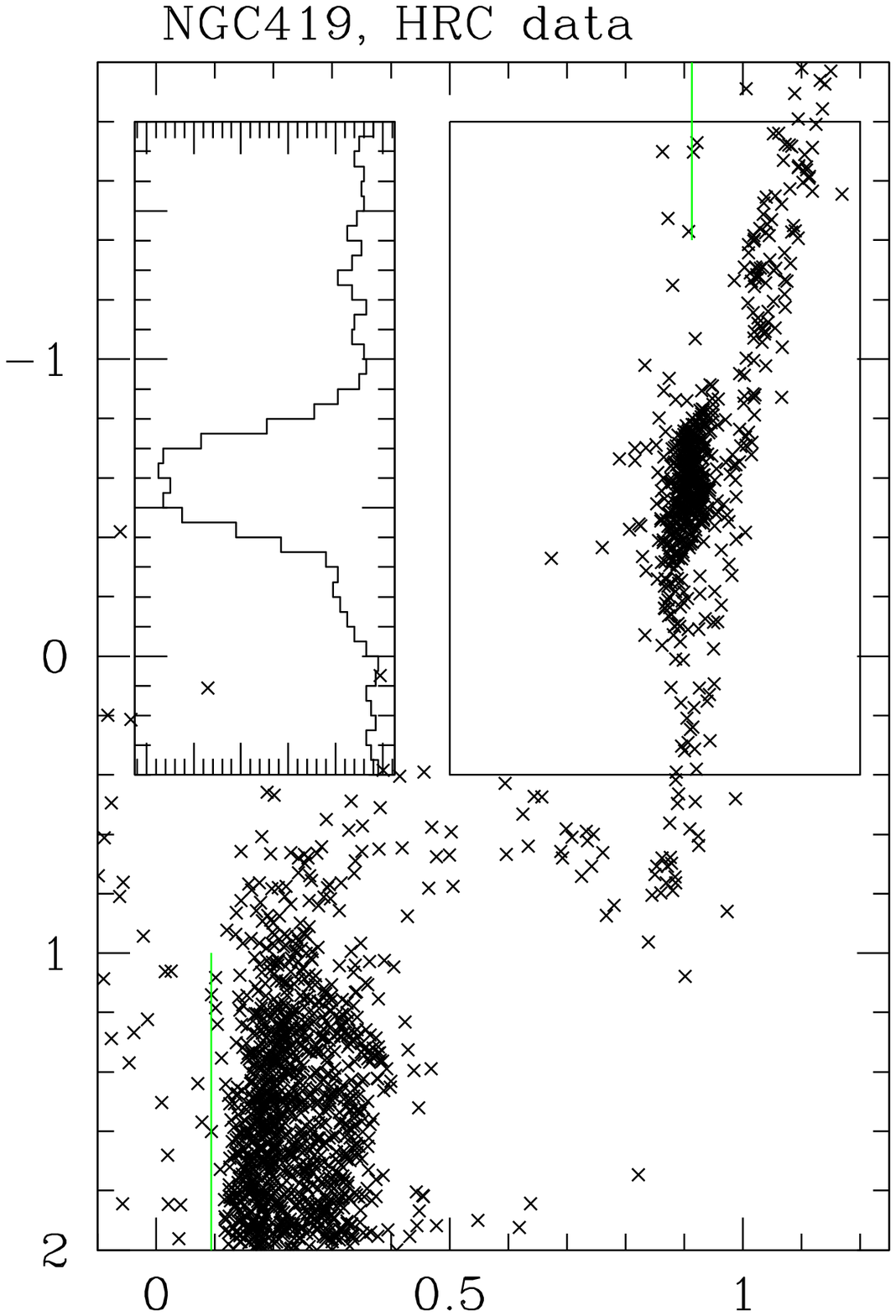}}
\end{minipage}
\\
\begin{minipage}{0.163\textwidth}
\resizebox{\hsize}{!}{\includegraphics{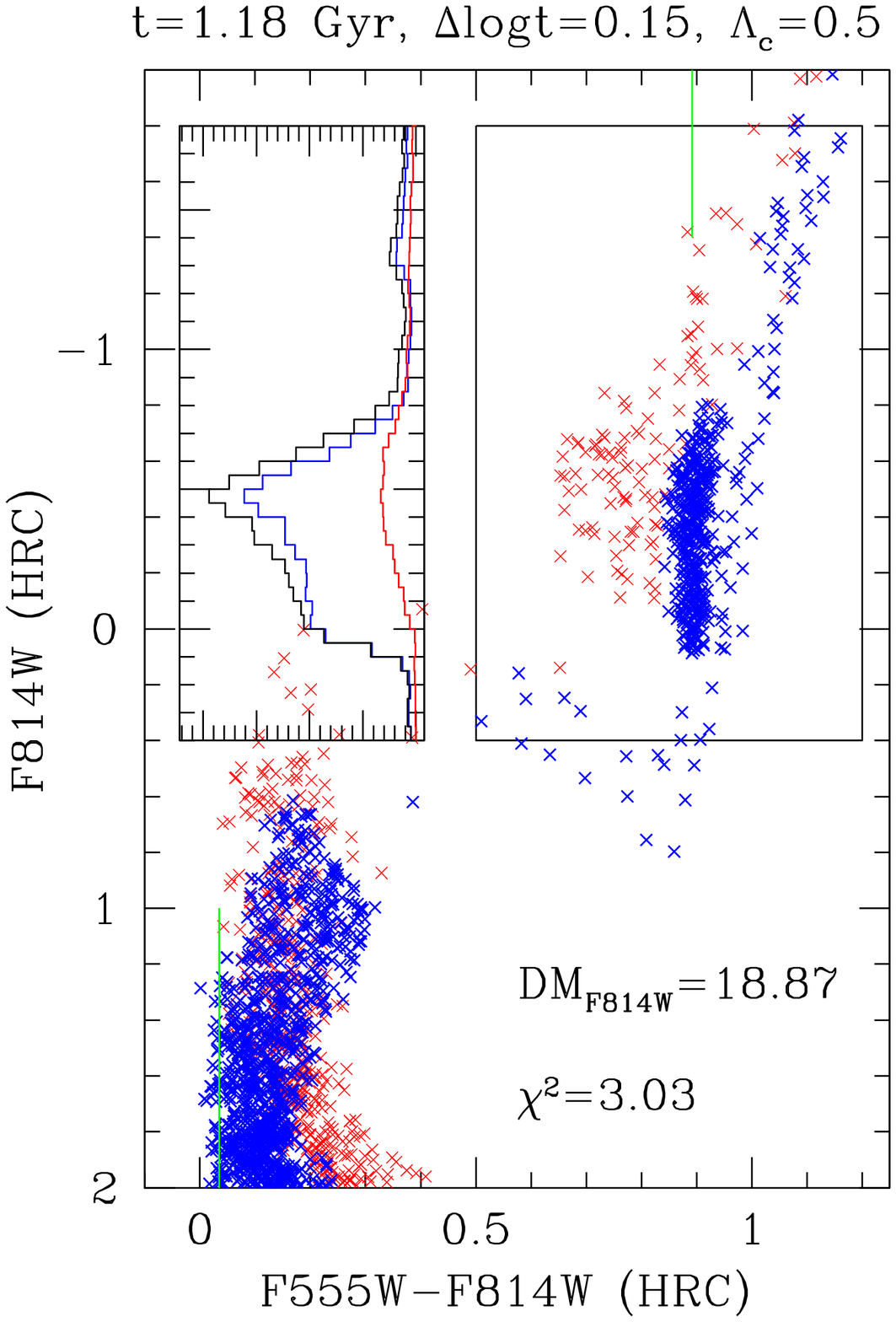}}
\end{minipage}
\begin{minipage}{0.163\textwidth}
\resizebox{\hsize}{!}{\includegraphics{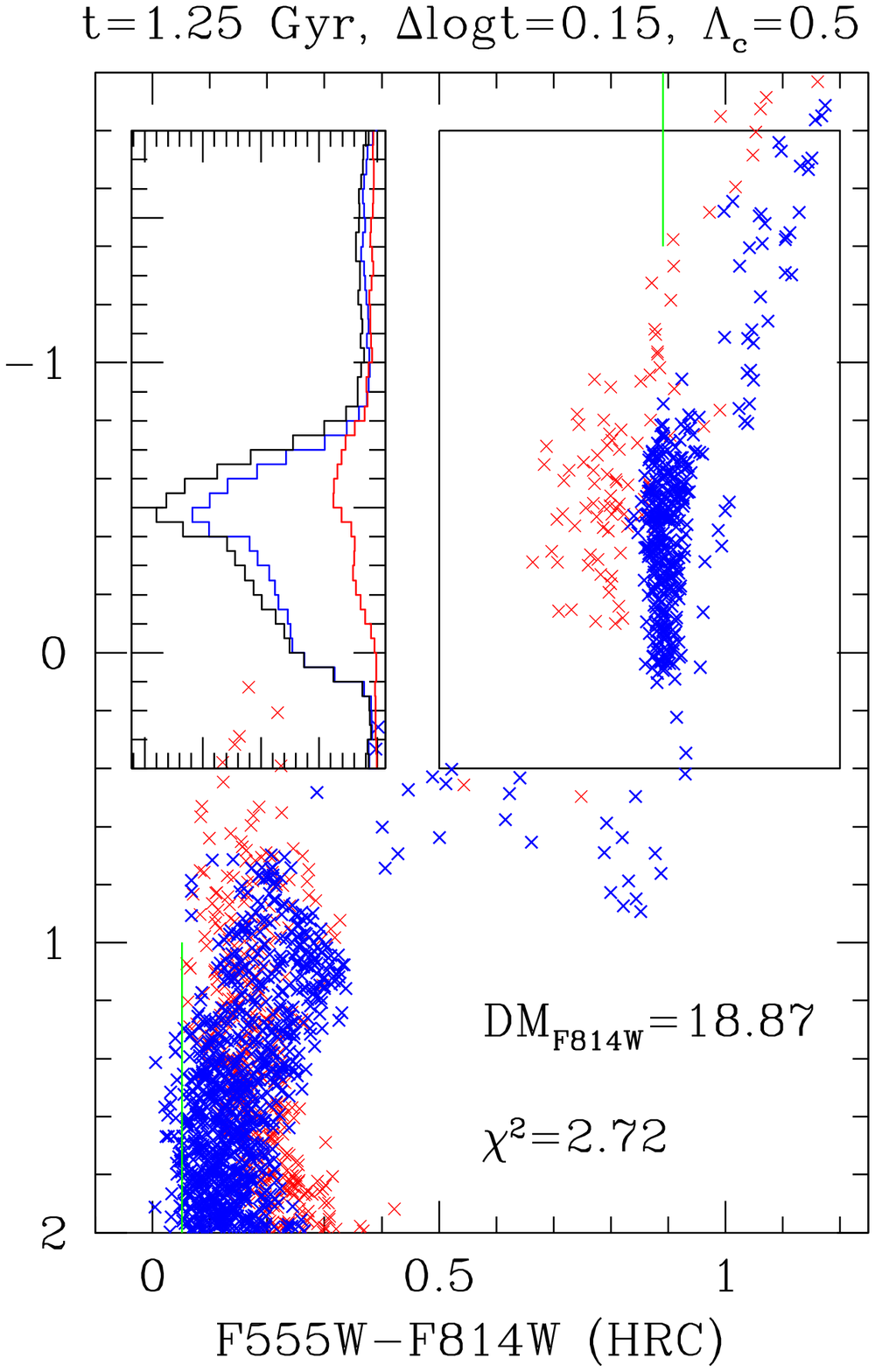}}
\end{minipage}
\begin{minipage}{0.163\textwidth}
\resizebox{\hsize}{!}{\includegraphics{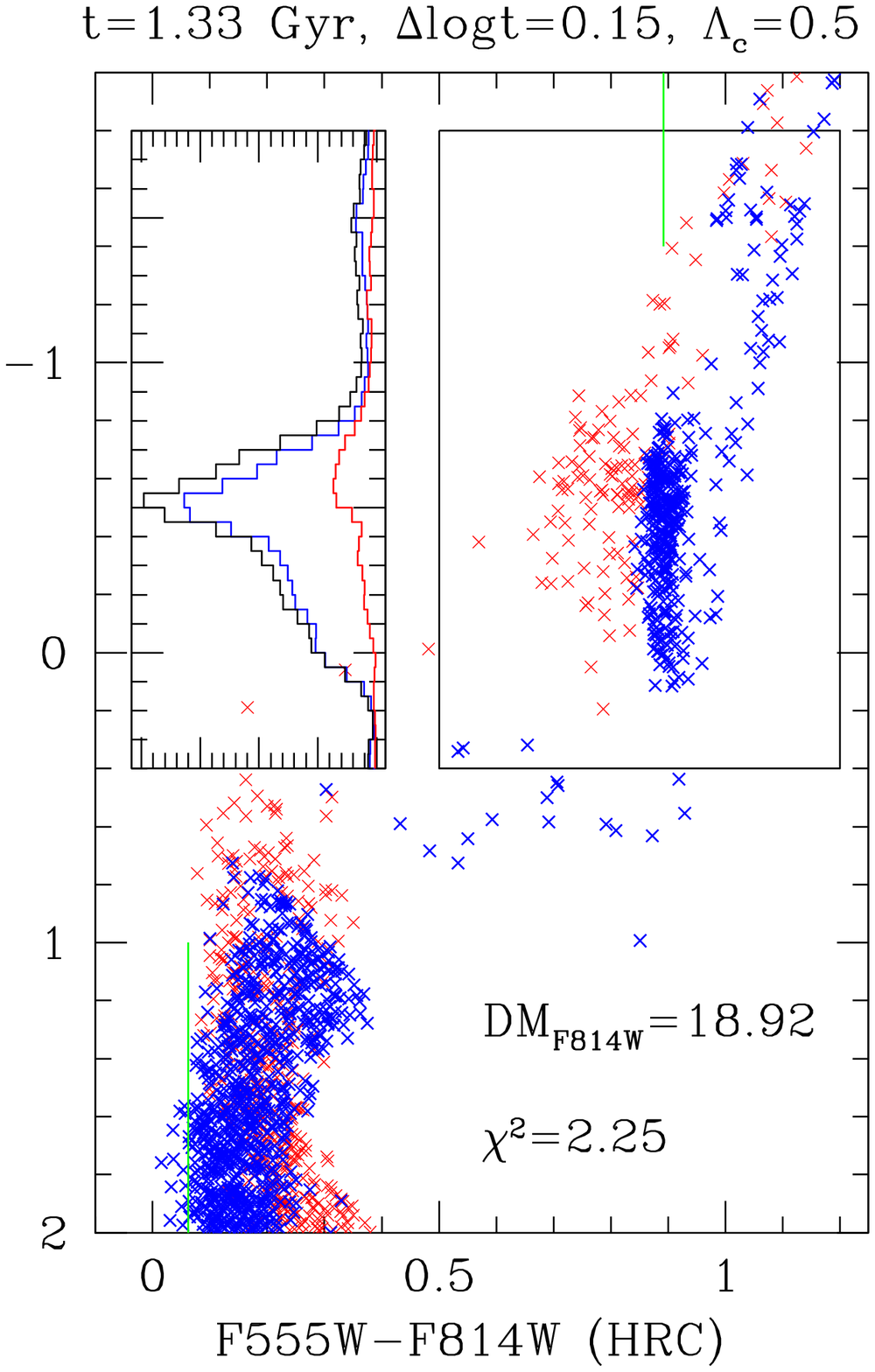}}
\end{minipage}
\begin{minipage}{0.163\textwidth}
\resizebox{\hsize}{!}{\includegraphics{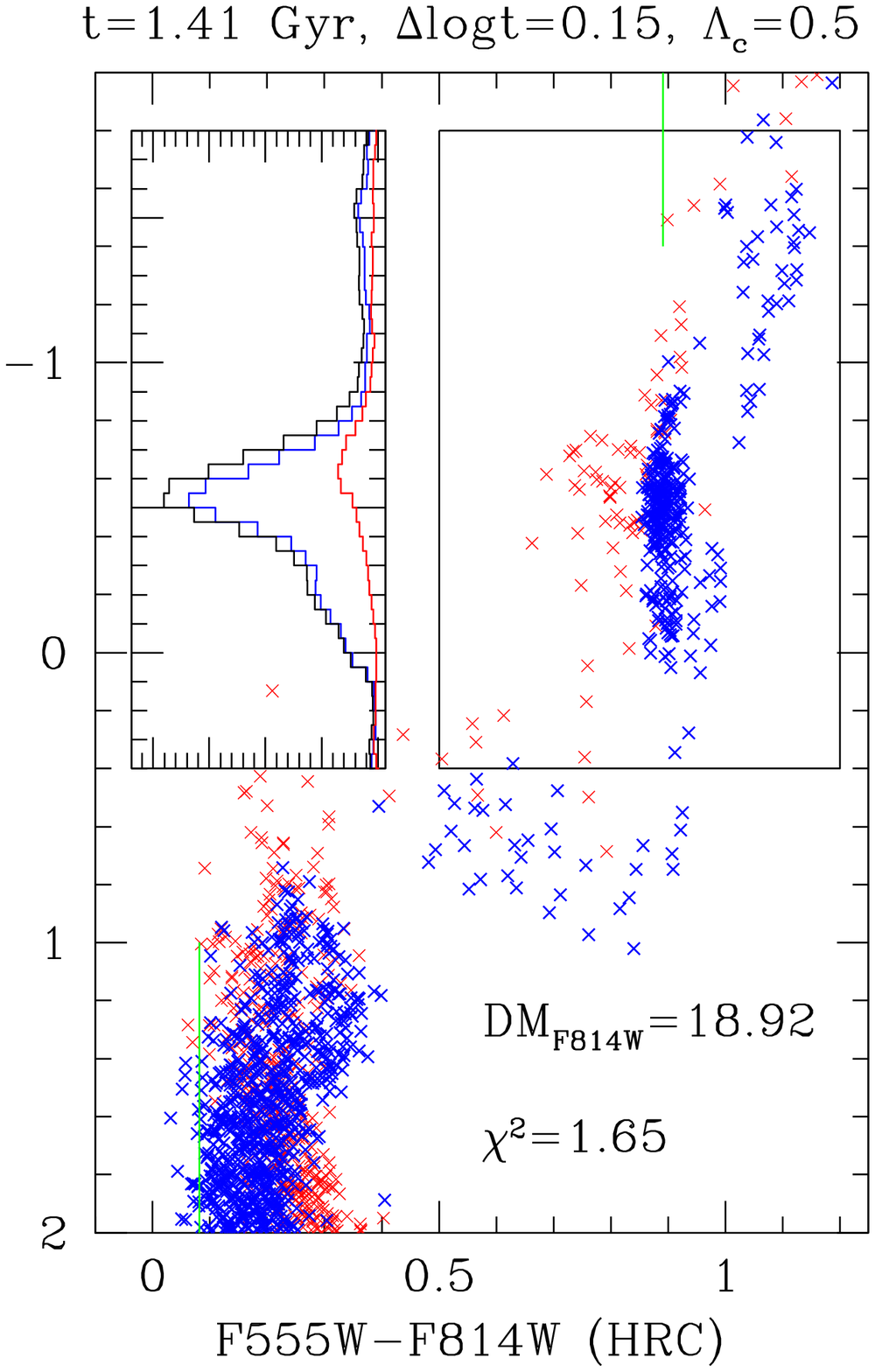}}
\end{minipage}
\begin{minipage}{0.163\textwidth}
\resizebox{\hsize}{!}{\includegraphics{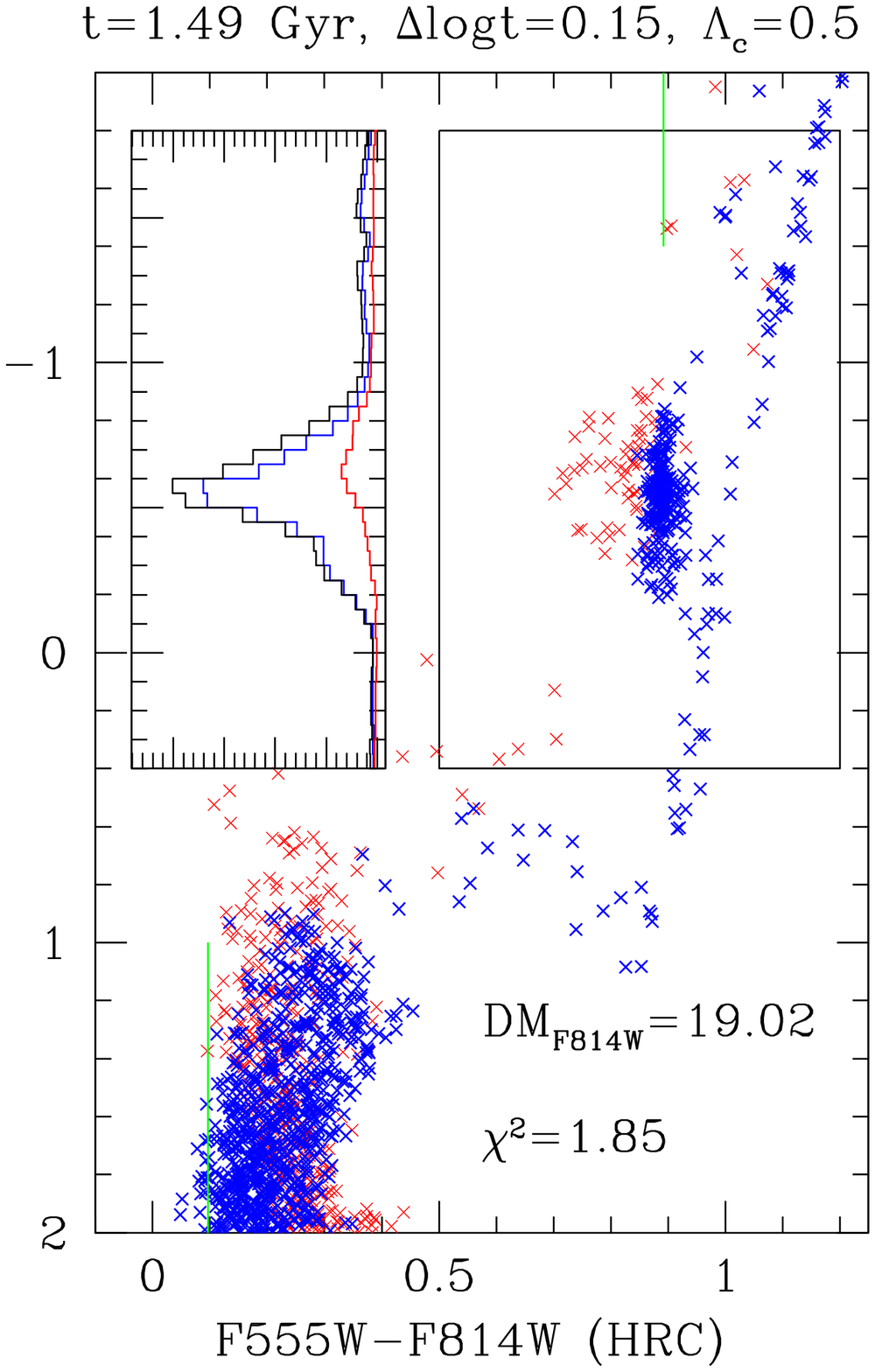}}
\end{minipage}
\begin{minipage}{0.163\textwidth}
\resizebox{\hsize}{!}{\includegraphics{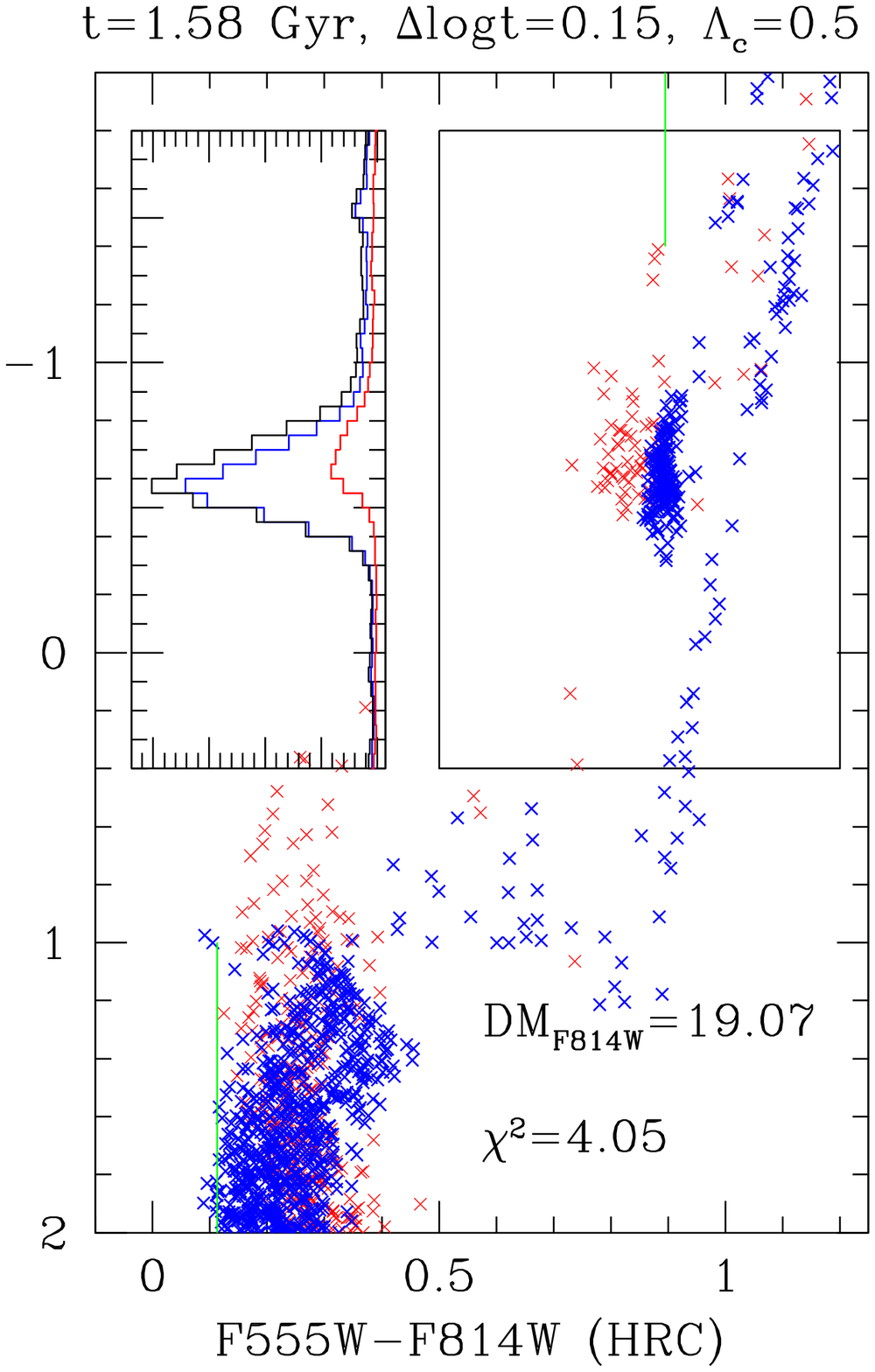}}
\end{minipage}
\caption{Models for the evolution of the red clump feature in the 
CMD as a function of mean population age $t$, for both a single-busrt
population (top panels) and for a composite one with duration of
$\Delta\log t\!=\!0.15$ (bottom panels), in both cases with the
assumption of moderate convective overshooting ($\Lambda_{\rm
c}\!=\!0.5$) and $Z\!=\!0.004$. Single stars are marked in blue,
double stars in red. Each panel presents on the top right a box
evincing the red clump, and on the top left the luminosity function
(LF) for the stars in this box. The best-fitting distance modulus and
the associated $\chi^2$
are also displayed.  For comparison, the top right panel shows the HRC
data of NGC~419 on the same scale, after being arbitrarily shifted by
19.0 and 0.09 in magnitude and colour, respectively. The green
vertical lines in all panels mark the median colour of the red clump,
and the bluest colour of the MS (see text). }
\label{fig_models}
\end{figure*}

\begin{figure*}
\begin{minipage}{0.163\textwidth}
\resizebox{\hsize}{!}{\includegraphics{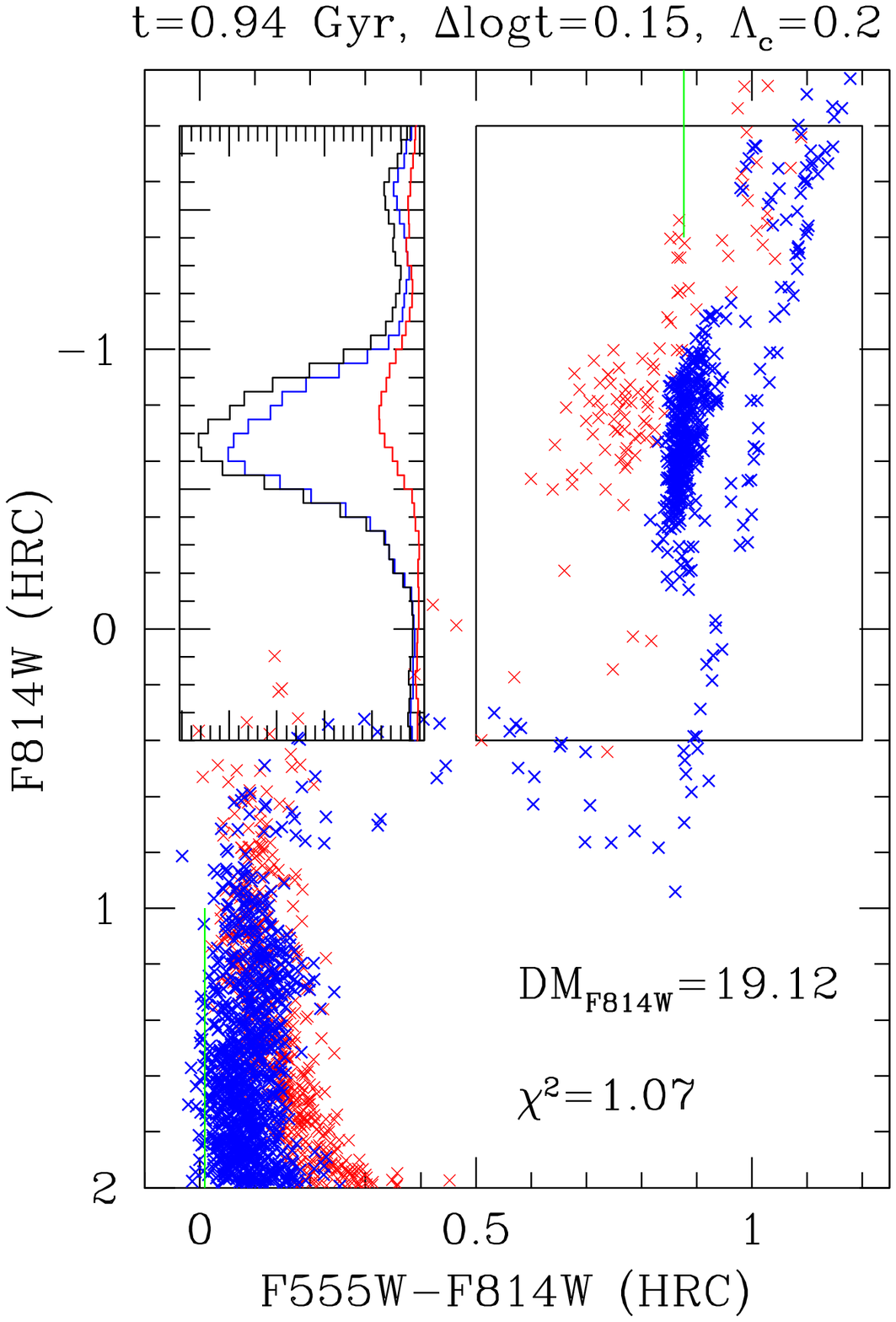}}
\end{minipage}
\begin{minipage}{0.163\textwidth}
\resizebox{\hsize}{!}{\includegraphics{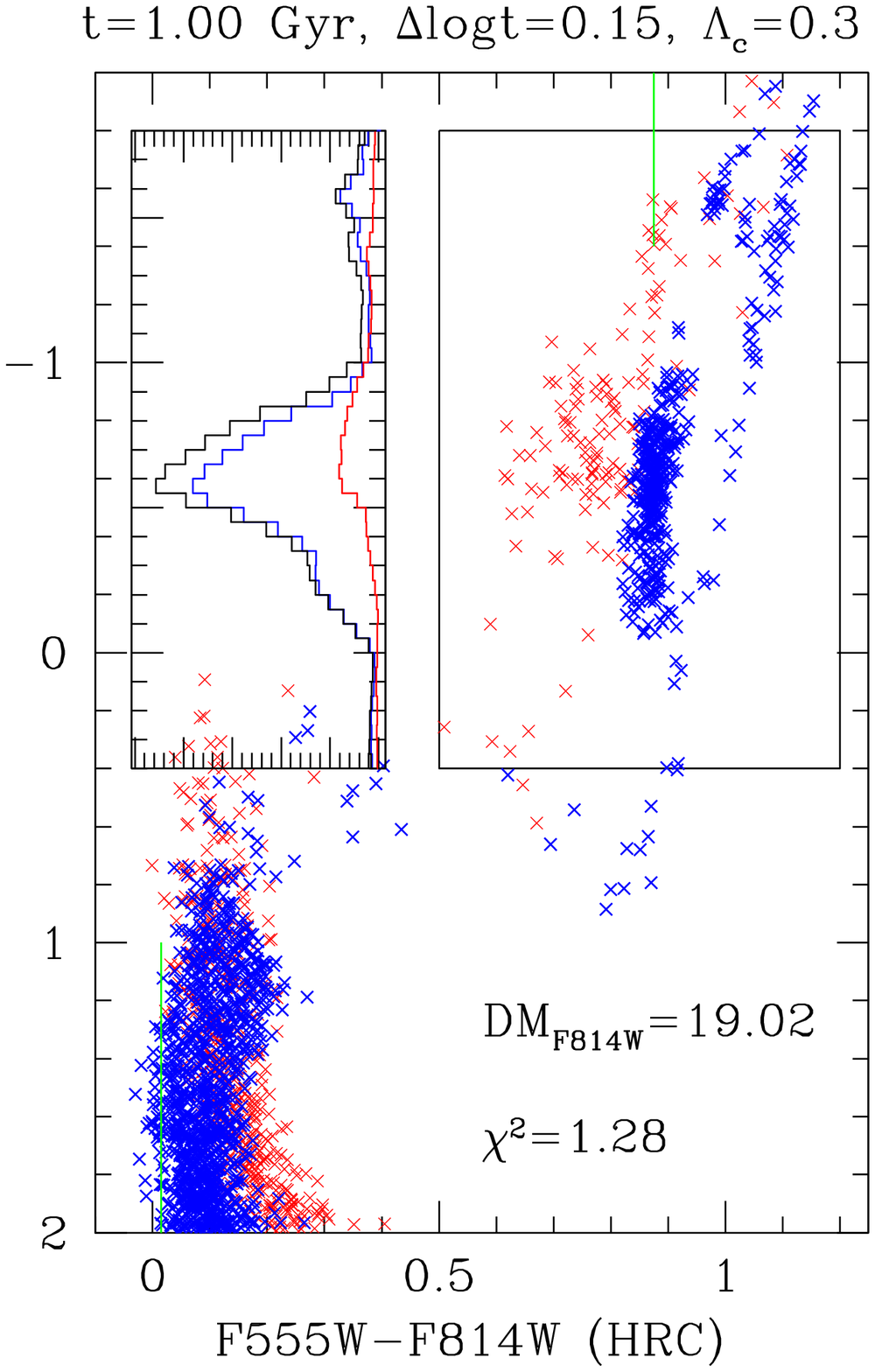}}
\end{minipage}
\begin{minipage}{0.163\textwidth}
\resizebox{\hsize}{!}{\includegraphics{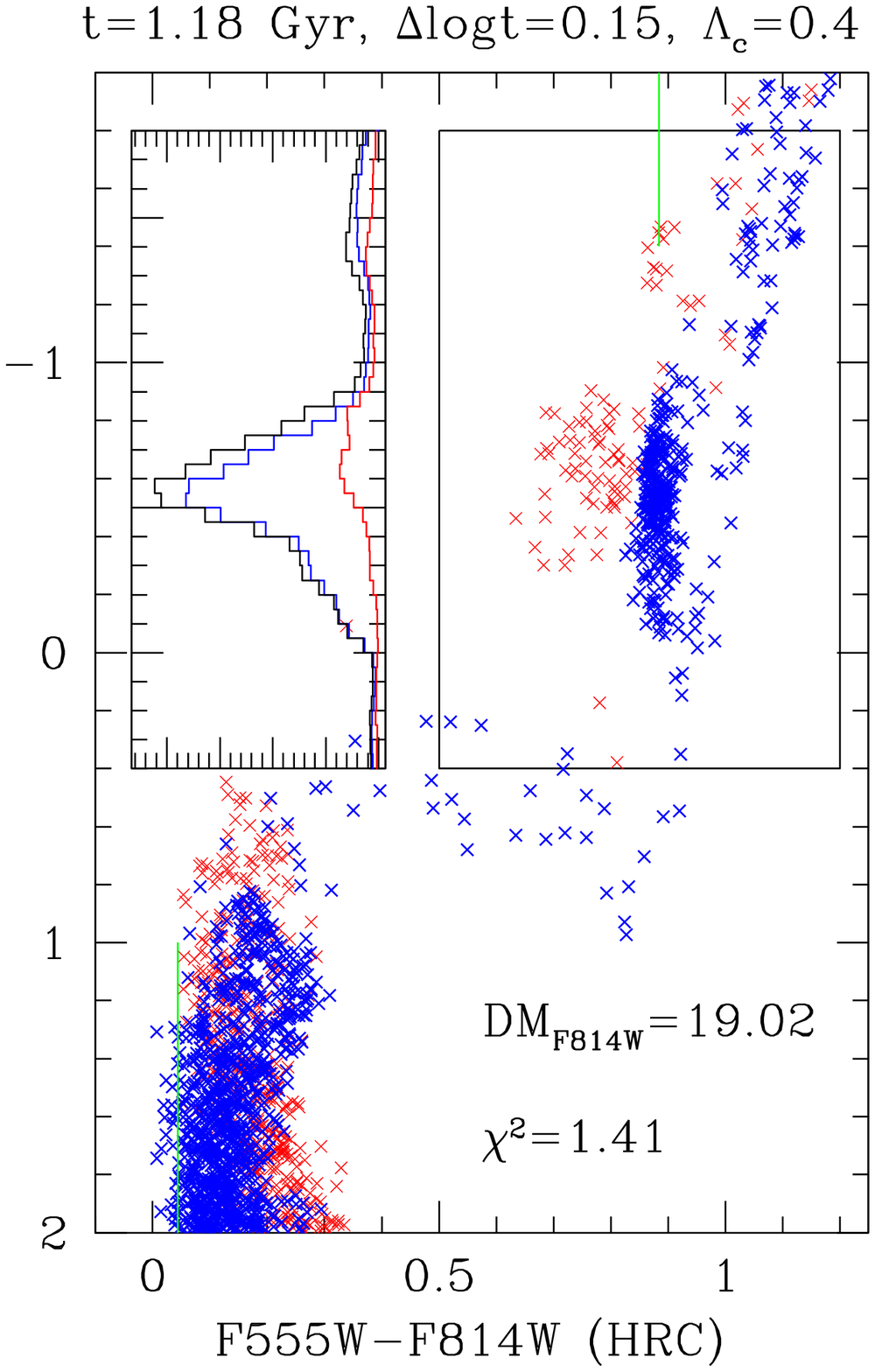}}
\end{minipage}
\begin{minipage}{0.163\textwidth}
\resizebox{\hsize}{!}{\includegraphics{out_lage9.15_d0.15_ov0.50_acs_hrc.ps}}
\end{minipage}
\begin{minipage}{0.163\textwidth}
\resizebox{\hsize}{!}{\includegraphics{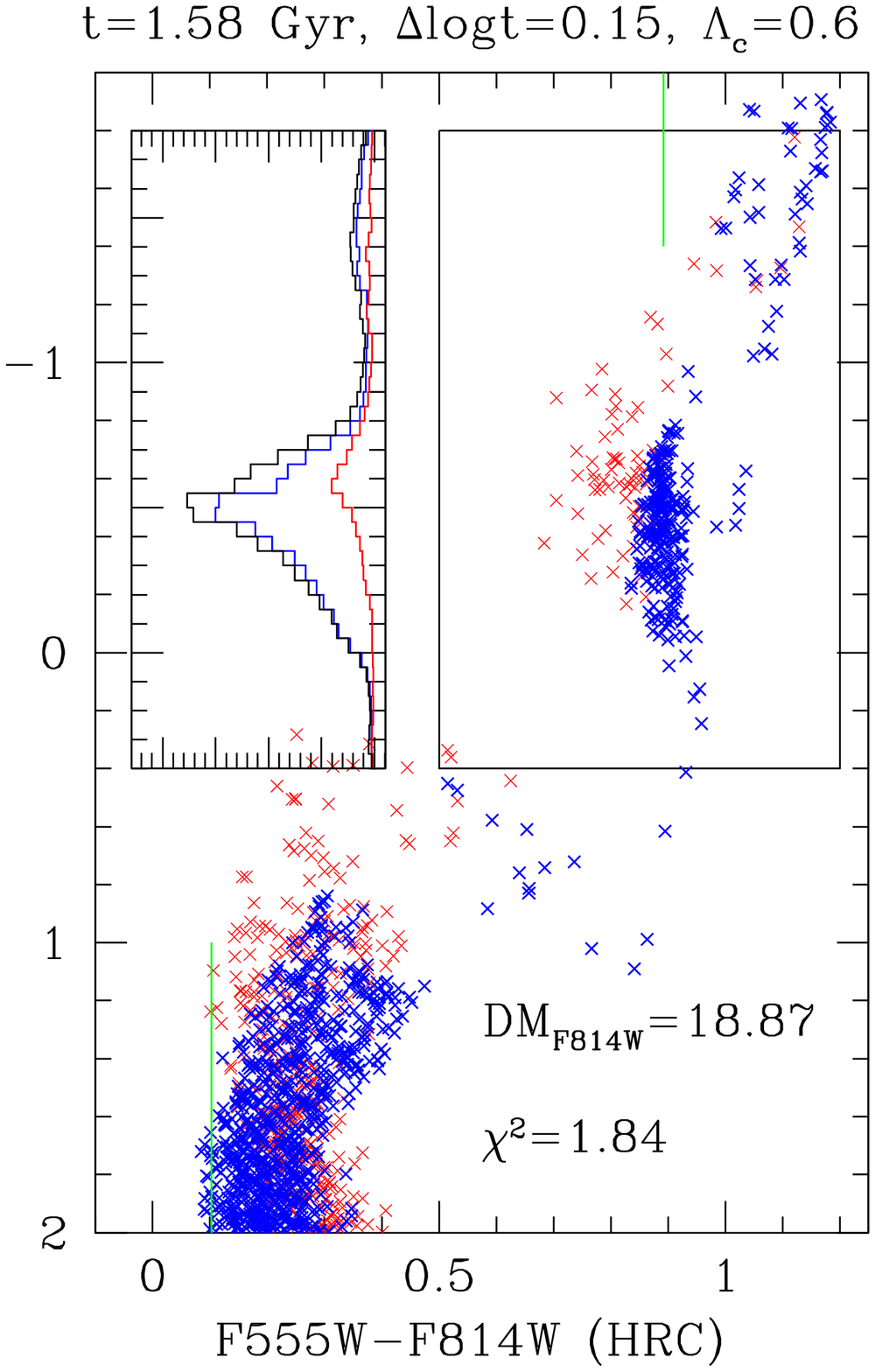}}
\end{minipage}
\begin{minipage}{0.163\textwidth}
\resizebox{\hsize}{!}{\includegraphics{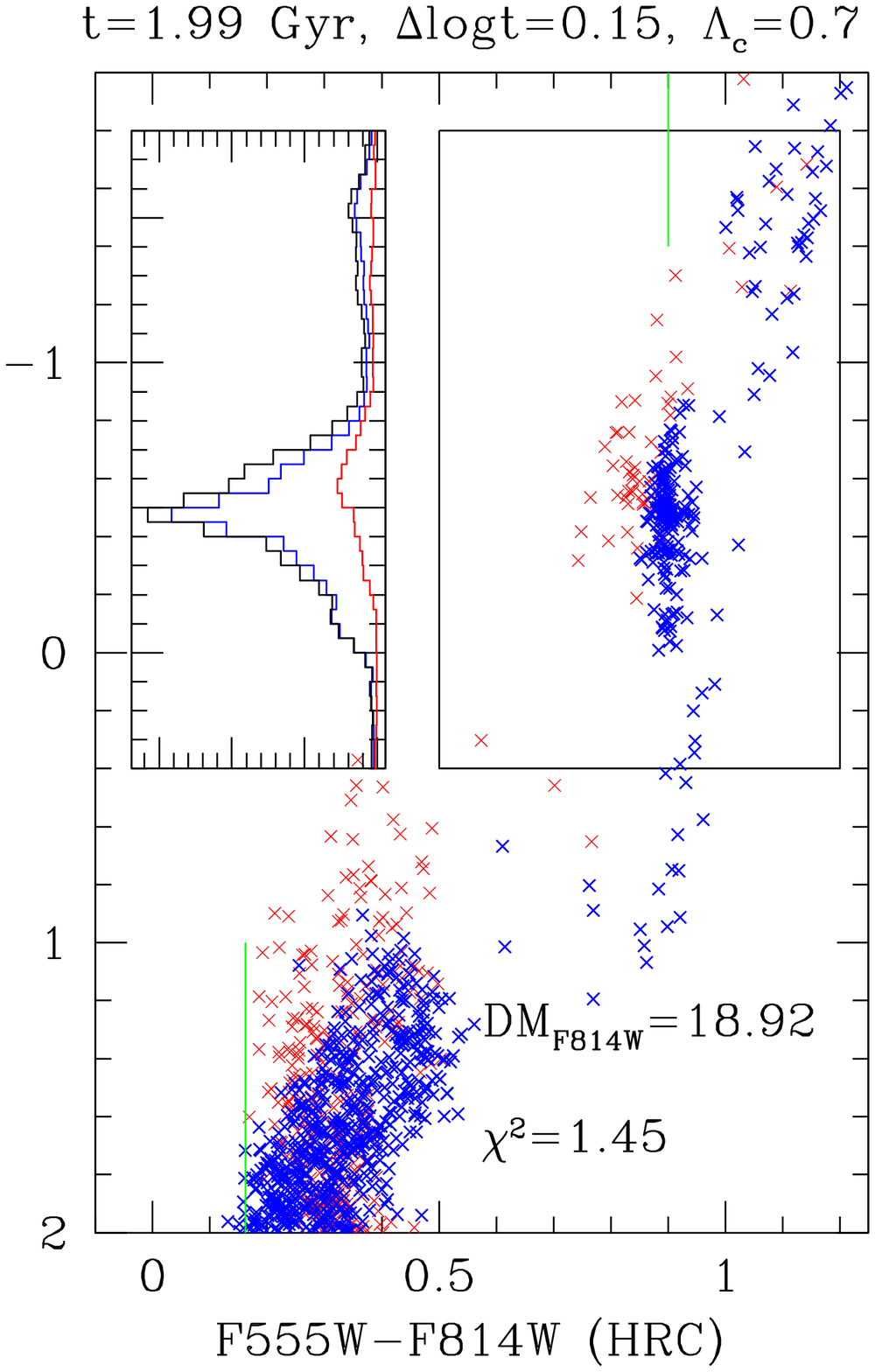}}
\end{minipage}
\caption{Top panels: The same as in Fig.~\ref{fig_models}, but now 
showing the models that best fit the red clump for several values of
overshooting efficiency $\Lambda_{\rm c}$, and for the $\Delta\log
t\!=\!0.15$ case only.  Although all models reproduce the observed red
clump similarly well, they differ very much in their age (from 0.94 to
1.99~Gyr, as $\Lambda_{\rm c}$ increases from 0.2 to 0.7), and produce
different MSTO magnitudes and colours. }
\label{fig_overshooting}
\end{figure*}

The isochrones are then fed to the TRILEGAL population synthesis code
\citep{Girardi_etal05} to simulate the photometry of star clusters at
the SMC distance. We apply to the TRILEGAL output the photometric
errors derived from artificial star tests performed on the original
HRC images. The results are illustrated in Fig.~\ref{fig_models},
which shows the expected time evolution of the red clump for a cluster
for two different cases: either for an almost-instantaneous burst of
star formation (with a duration of $\Delta\log t\!=\!0.01$), and for a
burst spanning a range of $\Delta\log t\!=\!0.15$. The latter case
corresponds roughly to the situation indicated by the MSTO stars in
Fig.~\ref{fig_cmd_hrc}. 20~\% of the stars in the simulation are
assumed to be detached binaries with a mass ratio comprised between
0.7 and 1
\citep{Woo_etal03}. Since this latter prescription is rather
uncertain, binaries are always marked with a different colour in our
plots.

It is evident from Fig.~\ref{fig_models} that the red clump rapidly
transits from a vertically-extended feature, to a much more compact
and slightly brighter clump at a mean age $t$ close to 1.25~Gyr. The
transition takes place completely in an age interval of just
$\sim\!0.1$~Gyr for the case of an intanstaneous burst, and in about
twice this time for the case of a $\Delta\log t\!=\!0.15$-wide
burst. From now on, we will refer to the mean age of this transition
as \thef.

We note that vertically-extended red clumps are present for all ages
younger than \thef, even in the intanstaneous-burst case; however,
they are wider than observed in NGC~419, and present drop-shaped LFs
-- i.e. with a sharp cut at the bottom and a more extended tail at the
top. This is not what observed in the red clump of NGC~419, which
presents the maximum of its LF at the top, together with a bump at the
faintest magnitudes. A configuration similar to NGC~419 is obtained
only in the extended-burst case, for ages comprised between 1.41 and
1.58~Gyr: indeed, this age range is the only one which combines the
vertically-extended red clump of younger ages, with the compact and
brighter red clump of older ages, in about the right proportions to
explain the observations. We identify NGC~419 as belonging to this
very limited -- and surely very singular -- age interval. In order to
identify the best-fitting age, we compute the $\chi^2$ between data
and model, for the red clump region only.
The distance modulus is varied until the minimum value of $\chi^2$ is
met for each age $t$. The results are printed in
Fig.~\ref{fig_models}, and evince the excellent quality of the fit for
the 1.41~Gyr-old model with $\Delta\log t\!=\!0.15$.

Finally, it is worth mentioning that the MS+red clump binaries tend to
draw a plume departing from the red clump towards bluer colours and
brighter magnitudes, which (1) is easily identifiable in observed CMDs
because of its colour separation from the red clump, and (2) do not
change the drop-shaped form of the LF for the younger red clumps.
Binaries cannot mimick the bimodal LF observed in the red clump of
NGC~419. 

A question raised by the referee is whether the two red clumps could
be caused by populations with different helium content, abundances of
CNO elements, or overshooting efficiency. Although nothing can be
excluded, so far there are no indications of such effects in clusters
as young as NGC~419. Moreover, it is Occam's razor to refrain us from
looking for more complicate alternatives: in fact, our explanation
requires {\em only} standard physics added to a quite simple
distribution of stellar ages -- the same one indicated by the
clusters' MSTO -- while keeping all other parameters constant. We
recall that the transition between faint and bright red clump is
something that inevitably happens, sooner or later, for {\em every}
single stellar population, causing {\em always} about the same amount
of brightening in the red clump over a similar timescale (which is
dictated mainly by the equation of state of partially-degenerate
matter). Therefore, there is no free parameter to be fixed in order to
explain the presence and position of the two red clumps, there is just
the very loose requirement of ``a prolonged-enough duration of the
star formation''.

\section{Overshooting and the age scale}
\label{constraints}

According to the interpretation given in this letter, the red clump in
NGC~419 corresponds to stars with a precise internal configuration
after H-exhaustion: their cores have a mass very close to 0.33~\Msun,
as their central temperatures approach $T_{\rm c}\!=\!10^8$~K. Slighty
higher core masses lead to non-degenerate He-ignition. Slightly
smaller core masses lead to e$^-$-degeneracy, which halts the core
contraction and is followed by the cooling of the central core by
plasma neutrinos; as a consequence the He-ignition is postponed to a
later stage -- namely the RGB tip -- at which the core masses have
grown up to 0.45~\Msun\ \citep[see][]{Sweigart_etal90}.

These core masses after H-exhaustion do also correspond to a narrow
interval of initial masses, comprehending the transition between
intermediate- and low-mass stars, \Mhef. It has long been known that
convective core overshooting changes the relation between the initial
mass and the H-exhausted core mass, hence directly affecting the value
of \Mhef, and its corresponding age \thef\
\citep[e.g.][]{Bressan_etal93}. Even if the efficiency $\Lambda_{\rm
c}$ can be constrained by means of several methods which use either
the morphology of the CMD or star count ratios \citep[e.g.][and
references therein]{Woo_etal03}, this is still a main source of
uncertainty in settling the age scale of intermediate-age clusters,
and in the theory of stellar populations in general.

Can the NGC~419 giants help us to set constraints on \Mhef\ and \thef,
and hence on $\Lambda_{\rm c}$?  Probably yes, considering that higher
\Mhef\ values (lower $\Lambda_{\rm c}$) imply bluer MSTOs. Once we
identify a star cluster during the particular age \thef, fixing the
position of its red clump(s) in the CMD, the relative position of the
MSTO should depend mainly on overshooting.

With this idea in mind, we have computed several set of stellar
evolutionary tracks and isochrones, with $\Lambda_{\rm c}$ varying
from 0.2 to 0.7. For each one of these sets we identify the age at
which, for a $\Delta\log t\!=\!0.15$ star formation burst, the red
clump morphology is best reproduced (just as in
Fig.~\ref{fig_models}). Indeed, Fig.~\ref{fig_overshooting} shows that
similarly good fits of the red clump morphology are obtained for all
values of $\Lambda_{\rm c}$, but at different ages $\thef$. We recall
that these different fits actually represent {\em very similar
distributions of the core mass after H-exhaustion}.

\begin{figure}
\resizebox{0.8\hsize}{!}{\includegraphics{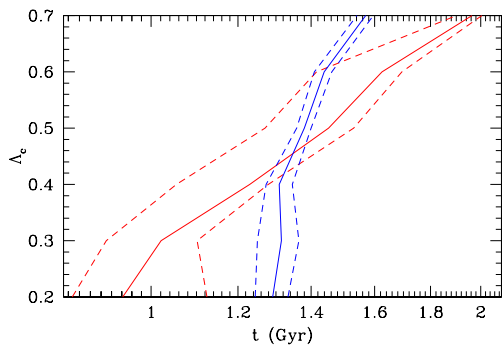}}
\caption{Confidence regions in the $t$ vs. 
$\Lambda_{\rm c}$ plane. The continuous red line follow the locus of
minimum $\chi^2$ values derived from the fitting of the red clump
morphology (see also Fig.~\ref{fig_overshooting}). The continuous blue
line describe the models which perfectly fit the colour difference
between the red clump and MSTO. Dashed lines present the estimated
70\% confidence limits.}
\label{fig_ages}
\end{figure}

We then measure the colour difference between the median of the red
clump, and the bluest border of the MSTO region defined by the 98\%
percentile of the star counts above an absolute magnitude of 2.5. This
quantity is little sensitive to the fraction of binaries, and our
simulations indicate that it can be measured with a $1\sigma$ error of
0.006~mag. Fig.~\ref{fig_ages} shows this quantity in the $t$
vs. $\Lambda_{\rm c}$ plane, together with the estimated 70\%
confidence level region of the best-fitting model to the red clump
morphology. The best simultaneous fit of the two quantities is
obtained for $\Lambda_{\rm c}\!=\!0.47_{-0.04}^{+0.14}$ and
$t=1.35_{-0.04}^{+0.11}$~Gyr (with random errors only).  

Importantly, these $t$ and $\Lambda_{\rm c}$ determinations are
largely free from uncertainties in the cluster distance and
reddening. However, they may be slightly affected by other factors
like the assumed fraction of binaries, the mixing-length parameter,
and the detailed star formation history of NGC~419. More detailed
analysis (in preparation) will aim to reduce these uncertainties by
using information from the complete CMD, and better exploring the
parameter space. A few preliminary conclusions can be advanced here:
(1) Models for $\Delta\log t$ values of 0.1 and 0.2 do also provide
good fits of the red clump morphology, with their best-fitting ages
differing by less than 6\% from those we find for $\Delta\log t=0.15$;
these age differences are comparable to the above-mentioned random
errors. However, such models have to be excluded because they clearly
provide a worst description of the MSTO region of the CMD. (2) On the
other hand, models with a fraction of binaries as small as 10\% tend
to provide slightly better fits of the red clump LF, at essentially
the same ages as those found with 20\% of binaries, and with just a
modest impact in the morphology of the MSTO.

\section{Final considerations}
\label{implications}

NGC~419 can be definitely added to the list of star clusters with a
secondary red clump, together with the Milky Way open clusters NGC~752
and 7789, and possibly also NGC~2660 and 2204, which were already
discussed by \citet{GirardiMermilliod_etal00}. This time, however, we
are facing a very populous cluster which presents a CMD rich of
details, from its lower MS up to the AGB carbon star sequence (see
Fig.~\ref{fig_cmd_hrc}, and \citealt{FMB90}). The secondary red clump
itself is very well populated and its detection can hardly be
controversial. This fine CMD feature provides strong constraints to
the core mass reached by its MS stars. All these aspects make of
NGC~419 an excellent tool for calibrating stellar evolution models, as
well as the age sequence of intermediate-age populations.

Can we identify additional star clusters in the Magellanic Clouds,
having the same secondary clump feature? Probably yes, since these
galaxies contain dozens of populous clusters with ages around 1~Gyr,
and the presence of multiple turn-offs is a common feature among them
\citep{Milone_etal08}. Indeed, from a rapid eye inspection of published 
CMDs obtained with HST/ACS, we notice that dual (main+secondary) red
clump structures seem to be present also in the LMC clusters NGC~1751,
1783, 1806, 1846, 1852, and 1917 -- see figs.~7, 8, 9, 16 and 17 in
\cite{Milone_etal08}, and fig.~1 in \cite{Mackey_etal08}. All 
these clusters have multiple turn-offs, with the youngest one being at
$F814W\!\sim\!19.5$, which is comparable with the NGC~419 one if we
consider the $\sim\!0.5$~mag difference in the SMC--LMC distance
moduli. A subsequent paper will examine these clusters in close
detail, in the perspective of deriving more stringent constraints to
stellar evolutionary models. Cluster fundamental parameters such as
the age, distance and reddening, will be re-evaluated as well.

\section*{Acknowledgments}
The data presented in this paper were obtained from the Multimission
Archive at the Space Telescope Science Institute (MAST). STScI is
operated by the Association of Universities for Research in Astronomy,
Inc., under NASA contract NAS5-26555.  We thank A. Bressan,
G. Bertelli and M. Clemens for the useful comments, and support from
INAF/PRIN07 CRA 1.06.10.03, contract ASI-INAF I/016/07/0, and the
Brazilian agencies CNPq and FAPESP.



\label{lastpage}


\begin{thebibliography}{}

\bibitem[\protect\citeauthoryear{{Bertelli}, {Girardi}, {Marigo} \&
  {Nasi}}{{Bertelli} et~al.}{2008}]{Bertelli_etal08}
{Bertelli} G.,  {Girardi} L.,  {Marigo} P.,    {Nasi} E.,  2008, \aap, 484, 815

\bibitem[\protect\citeauthoryear{{Bica}, {Geisler}, {Dottori}, {Clari{\'a}},
  {Piatti} \& {Santos} Jr.}{{Bica} et~al.}{1998}]{Bica_etal98}
{Bica} E.,  {Geisler} D.,  {Dottori} H.,  {Clari{\'a}} J.~J.,  {Piatti} A.~E.,
    {Santos} Jr. J.~F.~C.,  1998, \aj, 116, 723

\bibitem[\protect\citeauthoryear{{Bressan}, {Fagotto}, {Bertelli} \&
  {Chiosi}}{{Bressan} et~al.}{1993}]{Bressan_etal93}
{Bressan} A.,  {Fagotto} F.,  {Bertelli} G.,    {Chiosi} C.,  1993, \aaps, 100,
  647

\bibitem[\protect\citeauthoryear{{Bressan}, {Chiosi} \& {Bertelli}}{{Bressan}
  et~al.}{1981}]{Bressan_etal81}
{Bressan} A.~G.,  {Chiosi} C.,    {Bertelli} G.,  1981, \aap, 102, 25

\bibitem[\protect\citeauthoryear{{Frogel}, {Mould} \& {Blanco}}{{Frogel}
  et~al.}{1990}]{FMB90}
{Frogel} J.~A.,  {Mould} J.,    {Blanco} V.~M.,  1990, \apj, 352, 96

\bibitem[\protect\citeauthoryear{{Girardi}}{{Girardi}}{1999}]{Girardi99}
{Girardi} L.,  1999, \mnras, 308, 818

\bibitem[\protect\citeauthoryear{{Girardi}, {et~al.} }{{Girardi} et~al.}{2008}]{Girardi_etal08}
{Girardi} L., {et~al.},  2008,
  \pasp, 120, 583

\bibitem[\protect\citeauthoryear{{Girardi}, {Groenewegen}, {Hatziminaoglou} \&
  {da Costa}}{{Girardi} et~al.}{2005}]{Girardi_etal05}
{Girardi} L.,  {Groenewegen} M.~A.~T.,  {Hatziminaoglou} E.,    {da Costa} L.,
  2005, \aap, 436, 895

\bibitem[\protect\citeauthoryear{{Girardi}, {Groenewegen}, {Weiss} \&
  {Salaris}}{{Girardi} et~al.}{1998}]{Girardi_etal98}
{Girardi} L.,  {Groenewegen} M.~A.~T.,  {Weiss} A.,    {Salaris} M.,  1998,
  \mnras, 301, 149

\bibitem[\protect\citeauthoryear{{Girardi}, {Mermilliod} \&
  {Carraro}}{{Girardi} et~al.}{2000}]{GirardiMermilliod_etal00}
{Girardi} L.,  {Mermilliod} J.-C.,    {Carraro} G.,  2000, \aap, 354, 892

\bibitem[\protect\citeauthoryear{{Glatt}, {et~al.}}{{Glatt} et~al.}{2008}]{Glatt_etal08}
{Glatt} K.,  {et~al.},  2008, \aj, 136, 1703

\bibitem[\protect\citeauthoryear{{Grocholski} \& {Sarajedini}}{{Grocholski} \&
  {Sarajedini}}{2002}]{GrocholskiSarajedini02}
{Grocholski} A.~J.,  {Sarajedini} A.,  2002, \aj, 123, 1603

\bibitem[\protect\citeauthoryear{{Holtzman}, {et~al.}}{{Holtzman} et~al.}{1997}]{Holtzman_etal97}
{Holtzman} J.~A.,  {et~al.},  1997, \aj, 113, 656

\bibitem[\protect\citeauthoryear{{Mackey}, {Broby Nielsen}, {Ferguson} \&
  {Richardson}}{{Mackey} et~al.}{2008}]{Mackey_etal08}
{Mackey} A.~D.,  {Broby Nielsen} P.,  {Ferguson} A.~M.~N.,    {Richardson}
  J.~C.,  2008, \apjl, 681, L17

\bibitem[\protect\citeauthoryear{{Marigo}, {Girardi}, {Bressan}, {Groenewegen},
  {Silva} \& {Granato}}{{Marigo} et~al.}{2008}]{Marigo_etal08}
{Marigo} P.,  {Girardi} L.,  {Bressan} A.,  {Groenewegen} M.~A.~T.,  {Silva}
  L.,    {Granato} G.~L.,  2008, \aap, 482, 883

\bibitem[\protect\citeauthoryear{{Mermilliod}, {Mathieu}, {Latham} \&
  {Mayor}}{{Mermilliod} et~al.}{1998}]{Mermilliod_etal98}
{Mermilliod} J.-C.,  {Mathieu} R.~D.,  {Latham} D.~W.,    {Mayor} M.,  1998,
  \aap, 339, 423

\bibitem[\protect\citeauthoryear{{Milone}, {Bedin}, {Piotto} \&
  {Anderson}}{{Milone} et~al.}{2008}]{Milone_etal08}
{Milone} A.~P.,  {Bedin} L.~R.,  {Piotto} G.,    {Anderson} J.,  2008, arXiv:0810.2558

\bibitem[\protect\citeauthoryear{{Piatti}, {Geisler}, {Bica}, {Clari{\'a}},
  {Santos} Jr., {Sarajedini} \& {Dottori}}{{Piatti}
  et~al.}{1999}]{Piatti_etal99}
{Piatti} A.~E.,  {Geisler} D.,  {Bica} E.,  {Clari{\'a}} J.~J.,  {Santos} Jr.
  J.~F.~C.,  {Sarajedini} A.,    {Dottori} H.,  1999, \aj, 118, 2865

\bibitem[\protect\citeauthoryear{{Sirianni}, {et~al.}}{{Sirianni}
  et~al.}{2005}]{Sirianni_etal05}
{Sirianni} M.,  {et~al.},  2005, \pasp, 117, 1049

\bibitem[\protect\citeauthoryear{{Sweigart}, {Greggio} \& {Renzini}}{{Sweigart}
  et~al.}{1990}]{Sweigart_etal90}
{Sweigart} A.~V.,  {Greggio} L.,    {Renzini} A.,  1990, \apj, 364, 527

\bibitem[\protect\citeauthoryear{{Woo}, {Gallart}, {Demarque}, {Yi} \&
  {Zoccali}}{{Woo} et~al.}{2003}]{Woo_etal03}
{Woo} J.-H.,  {Gallart} C.,  {Demarque} P.,  {Yi} S.,    {Zoccali} M.,  2003,
  \aj, 125, 754

\end{thebibliography}
\end{document}